\documentclass[12pt]{article}
\usepackage{graphicx}
\usepackage{graphics}
\usepackage{amsthm}
\usepackage{amssymb, amsmath}
\title{Certain logarithmic integrals, including solution of Monthly problem \#tbd, zeta
values, and expressions for the Stieltjes constants} 
\author{Mark W. Coffey\\
Department of Physics\\
Colorado School of Mines\\
Golden, CO  80401\\
(Received $\mbox{~~~~~~~~~~~~~~~~~~~~~~~~~~~~~~~2011}$)}
\date{December 31, 2011}
\begin{document}
\maketitle
\baselineskip=25 pt
\begin{abstract}

We solve problem x proposed by O. Oloa, AMM xxx 2012 {\bf 119?} (to appear), p. yyy for 
certain definite logarithmic integrals.  A number of generating functions are developed with
certain coefficients $p_n$, and some extensions are presented.  The explicit relation of
$p_n$ to N\"{o}rlund numbers $B_n^{(n)}$ is discussed.  Certain inequalities are
conjectured for the $\{p_n\}$ sequence of coefficients, including its convexity, and an
upper bound is demonstrated.  It is shown that $p_n$ values may be used
to express the Stieltjes constants for the Hurwitz and Riemann zeta functions, as well as
values of these zeta functions at integer argument.  Other
summations with the $p_n$ coefficients are presented.

\end{abstract}
 
\medskip
\baselineskip=15pt
\centerline{\bf Key words and phrases}
\medskip 

\noindent

logarithmic integrals, Pochhammer symbol, generating function, digamma function, Glaisher
constant, N\"{o}rlund number, Hurwitz zeta function, Stieltjes constants 

\vfill
\centerline{\bf 2010 AMS codes} 
11Y60, 11Y35, 05A15

\baselineskip=25pt
\pagebreak
\medskip
\centerline{\bf Solution of problem xxx}
\medskip

O. Oloa has proposed the following problem in the Amer. Math. Monthly {\bf 119} (?), 
yyy (2012?).  

\noindent
(a) Prove the formula
$$\int_0^1\left({1 \over {\ln x}}+{1 \over {1-x}}\right)^2dx=\ln(2\pi)-{3 \over 2}. \eqno(1.1)$$
(b) If $\sigma \geq 0$, find a closed form expression for
$$\int_0^1\left({1 \over {\ln x}}+{1 \over {1-x}}\right)^2 x^\sigma dx.$$

N.B. Although the statement (b) is restricted to $\sigma \geq 0$, this part may be
extended to Re $\sigma >-1$, and our explicit expressions reflect this fact.

We first prove (1.1) and answer part (b), and then present several extensions.

{\it Proof}.  Let $(a)_n=\Gamma(a+n)/\Gamma(a)$ be the Pochhammer symbol, where $\Gamma$
is the Gamma function, $\psi=\Gamma'/\Gamma$ be the digamma function, $\gamma=-\psi(1)$ be the Euler constant, and $_pF_q$ the generalized hypergeometric function (e.g., \cite{grad,nbs}). 
We introduce the positive constants (e.g., \cite{coffey2009}, Proposition 11, \cite{coffeyjnt}, Proposition 5, \cite{coffeyseries}, Proposition 2)
$$p_{n+1}=-{1 \over {n!}}\int_0^1 (-x)_n dx={{(-1)^{n+1}} \over {n!}}\sum_{k=1}^n
{{s(n,k)} \over {k+1}}, \eqno(1.2)$$
where $s(k,\ell)$ is the Stirling number of the first kind.  The first few values
of these are $p_2=1/2$, $p_3=1/12$, $p_4=1/24$, $p_5=19/720$, and $p_6=3/160$.  These constants enter the generating function 
$$\sum_{n=1}^\infty p_{n+1}z^{n-1} ={1 \over z}+{1 \over {\ln(1-z)}}, ~~~~~~|z|<1.  \eqno(1.3)$$
Multiplying (1.3) by $\ln(1-z)$ and manipulating series, one finds the recursion relation
$$p_{n+1}={1 \over {n+1}}-\sum_{j=1}^{n-1} {p_{j+1} \over {(n-j+1)}}, ~~~~n \geq 1.$$  
We then have
\newline{\bf Lemma 1}.
$${1 \over {\ln^2(1-z)}}=\sum_{n=1}^\infty [(n+1)p_{n+3}-np_{n+2}]z^n+p_3+{{(1-z)} \over z^2},
~~~~~~|z|<1.  \eqno(1.4)$$

This follows from the derivative expression
$$\sum_{n=2}^\infty (n-1)p_{n+1}z^{n-2}=-{1 \over z^2}+{1 \over {(1-z)\ln^2(1-z)}}.  \eqno(1.5)$$

As a consequence, we obtain
$$\left({1 \over {\ln(1-z)}}+{1 \over z}\right)^2={1 \over 4}+\sum_{n=1}^\infty [(n+3)p_{n+3}-np_{n+2}]z^n.  \eqno(1.6)$$

Then for (a),
$$\int_0^1 \left({1 \over {\ln(1-z)}}+{1 \over z}\right)^2dz=\int_0^1\left({1 \over {\ln z}}+{1 \over {1-z}}\right)^2dz$$
$$=\int_0^1\left\{{1 \over 4}+\sum_{n=1}^\infty [(n+3)p_{n+3}-np_{n+2}](1-z)^n \right\} dz$$
$$={1 \over 4}+\sum_{n=1}^\infty [(n+3)p_{n+3}-np_{n+2}]{1 \over {n+1}}.  \eqno(1.7)$$
The integral representation of (1.2) may now be inserted, and the sums rewritten in
terms of binomial coefficients.  For instance, we have 
$$\sum_{n=1}^\infty {1 \over {n+1}}{{(n+3)} \over {(n+2)!}}(-x)_{n+2}
=\sum_{n=1}^\infty (-1)^n {{(n+3)} \over {(n+1)}} {x \choose {n+2}}=\sum_{n=1}^\infty (-1)^n \left[1+{2 \over {(n+1)}}\right] {x \choose {n+2}}$$
$$={1 \over 2}(2-x)(x-1)-x[1-2\gamma+x-2\psi(x+1)].  \eqno(1.8)$$
Noting that $\int_0^1 \psi(x+1)dx=0$, we have
$$\int_0^1 \left({1 \over {\ln(1-z)}}+{1 \over z}\right)^2dz={1 \over 4}+
\int_0^1\left[\gamma-{x \over 2}-2\gamma x+{3 \over 2} x^2-2x\psi(x+1)\right]dx$$
$$=\ln(2\pi)-{3 \over 2}.  \eqno(1.9)$$
In the last step, we integrated by parts (e.g., \cite{loggamma}),
$$\int_0^1 x \psi(x+1)dx=-\int_0^1 \ln \Gamma(x+1)dx=-\int_0^1 [\ln x+\ln \Gamma(x)]dx
=1-{1 \over 2}\ln(2\pi).  \eqno(1.10)$$


For (b), we let $B(x,y)=\Gamma(x)\Gamma(y)/\Gamma(x+y)$ be the Beta function.  Then
$$\int_0^1\left({1 \over {\ln x}}+{1 \over {1-x}}\right)^2 x^\sigma dx$$
$$=\int_0^1\left\{{1 \over 4}+\sum_{n=1}^\infty [(n+3)p_{n+3}-np_{n+2}](1-x)^n \right\}x^\sigma dx$$
$$={1 \over {4(\sigma+1)}}+\sum_{n=1}^\infty [(n+3)p_{n+3}-np_{n+2}]B(n+1,\sigma+1). \eqno(1.11)$$
\noindent
{\bf Lemma 2}.  For Re $y>0$,
$$\sum_{n=1}^\infty p_{n+3}B(n+1,y)=-{1 \over 2}\left[{1 \over {6y}}-1+2y-\ln(2\pi)+2\ln \Gamma(y)
+(1-2y)\psi(y)\right].  \eqno(1.12)$$

{\it Proof}.  From (1.2) we have
$$\sum_{n=1}^\infty p_{n+3}B(n+1,y)=-{1 \over {2y}}\int_0^1 x(x-1)[~_3F_2(1,1,2-x;3,y+1;1)-1]dx.
\eqno(1.13)$$
We now use the identity
$${{(1)_j} \over {(3)_j}}=2{{(1)_j} \over {(2)_j}}-{{(2)_j} \over {(3)_j}}, \eqno(1.14)$$
being a special case of
$${{(a)_j} \over {(a+2)_j}}=(a+1){{(a)_j} \over {(a+1)_j}}-a{{(a+1)_j} \over {(a+2)_j}},$$
to obtain
$$~_3F_2(1,1,2-x;3,y+1;1)=\sum_{j=0}^\infty {{(1)_j} \over {(3)_j}}{{(2-x)_j} \over {(y+1)_j}}$$
$$=\sum_{j=0}^\infty \left[2{{(1)_j} \over {(2)_j}}-{{(2)_j} \over {(3)_j}}\right]{{(2-x)_j} \over {(y+1)_j}}{{(1)_j} \over {j!}}$$
$$={{2y} \over {x(x-1)}}[1-x-(x+y-1)\psi(y)+(x+y-1)\psi(x+y+1)].  \eqno(1.15)$$
Carrying out the integration of (1.13) gives the Lemma.  \qed

Then by Proposition 2 in the next section with $y=\sigma+1$ we obtain
$$\int_0^1\left({1 \over {\ln x}}+{1 \over {1-x}}\right)^2 x^\sigma dx
=(\sigma+1)\ln(\sigma+1)-2\sigma+\sigma \psi(\sigma+1)-2\ln\Gamma(\sigma+1)+\ln(2\pi)-{3 \over
2}.  \eqno(1.16)$$

{\it Remarks}.  As is apparent from (1.11) and (1.16), Re $\sigma=-1$ is the `critical line' 
for divergence of the integral.

We have the following
{\newline \bf Corollary 1}.  For $n \geq 1$,
$$(n+3)p_{n+3}-np_{n+2}=\sum_{k=1}^{n+1} p_{k+1}p_{n-k+3}.  \eqno(1.17)$$

This follows from multiplication of series, using (1.6),
$$\left({1 \over {\ln(1-z)}}+{1 \over z}\right)^2={1 \over 4}+\sum_{n=1}^\infty [(n+3)p_{n+3}-np_{n+2}]z^n$$
$$=\sum_{n=0}^\infty \sum_{k=1}^{n+1} p_{k+1}p_{n-k+3}z^n.  \eqno(1.18)$$
Is there a combinatorial interpretation of identity (1.13)?

The Appendix generalizes (1.14) and the following identity for ratios of Pochhammer symbols.

\medskip
\centerline{\bf Extensions}
\medskip

We may proceed similarly as above, and find for instance
$${1 \over {\ln^3(1-z)}}={1 \over 2}\sum_{n=1}^\infty [(n+1)(n+2)p_{n+4}-(n+1)(2n+1)p_{n+3}+n^2 p_{n+2}]z^n$$
$$-{1 \over z^3}+{3 \over {2z^2}}-{1 \over {2z}}, ~~~~~~|z|<1, \eqno(2.1)$$
giving, along with (1.3) and (1.4),
$$\left({1 \over {\ln(1-z)}}+{1 \over z}\right)^3={1 \over {\ln^3(1-z)}}+{3 \over {z\ln^2(1-z)}}+{3 \over {z^2\ln(1-z)}}+{1 \over z^3}$$
$$={1 \over 2}\sum_{n=4}^\infty[(n+4)(n+5)p_{n+4}-(n+1)(2n+7)p_{n+3}+n^2p_{n+2}]z^n$$
$$+{1 \over 8}+{z \over {16}}-{z^2 \over {24}}+{{133} \over {4320}}z^3.  \eqno(2.2)$$

Then
$$\int_0^1 \left({1 \over {\ln(1-z)}}+{1 \over z}\right)^3dz=\int_0^1\left({1 \over {\ln z}}
+{1 \over {1-z}}\right)^3dz$$
$$={1 \over 2}\sum_{n=4}^\infty[(n+4)(n+5)p_{n+4}-(n+1)(2n+7)p_{n+3}+n^2p_{n+2}]{1 \over {(n+1)}} + {{3073} \over {17280}}$$
$$=-{{31} \over {24}}+6 \ln A,  \eqno(2.3)$$
wherein $A$ is Glaisher's constant, such that $\ln A=-[\zeta(-1)+\zeta'(-1)]=1/12-\zeta'(-1)$,
and $\zeta(s)$ is the Riemann zeta function.  The latter contribution enters from the
integral
$$\int_0^1 t^2  \psi(t+1)dt=-2\int_0^1 t\ln \Gamma(t+1)dt={1 \over 2}(1-\ln 2\pi)+2\ln A.   
\eqno(2.4)$$
This integral may be readily determined from Kummer's Fourier series for $\ln \Gamma$.
Otherwise, it may be found through the infinite series
$$\int_0^1 t^2 \psi(t+1)dt=-{\gamma \over 3}+\sum_{k=2}^\infty (-1)^k \zeta(k) \int_0^1 t^{k+1}
dt=-{\gamma \over 3}+\sum_{k=2}^\infty {{(-1)^k} \over {k+2}} \zeta(k).  \eqno(2.5)$$

For reference, we have from (2.1)
$${1 \over {\ln^4(1-z)}}={1 \over 6}\sum_{n=1}^\infty \left[(n+1)(n+2)(n+3)p_{n+5} -3(n+1)^2(n+2)p_{n+4} \right.$$
$$\left. +(n+1)(3n^2+3n+1)p_{n+3}-n^3 p_{n+2}\right]z^n -{1 \over {720}}-{1 \over {6z}}
+{7 \over {6z^2}}-{2 \over z^3}+{1 \over z^4}, ~~~~~~|z|<1. \eqno(2.6)$$

Then also using (1.3), (1.4), and (2.1), we find
$$\left({1 \over {\ln(1-z)}}+{1 \over z}\right)^4$$
$$={1 \over 6}\sum_{n=1}^\infty[(n+1)(n+2)
(n+3)p_{n+5}-3(n+1)^2(n+2)p_{n+4}+(n+1)(3n^2+3n+1)p_{n+3}-n^3p_{n+2}]z^n$$
$$-{1 \over {720}}+2\sum_{n=1}^\infty[(n+1)(n+2)p_{n+4}-(n+1)(2n+1)p_{n+3}+n^2p_{n+2}]
z^{n-1}$$
$$+6\sum_{n=1}^\infty[(n+2)p_{n+4}-(n+1)p_{n+3}]z^{n-1}+4\sum_{n=1}^\infty p_{n+4}z^{n-1}.
\eqno(2.7)$$
By using (1.2) we obtain
$$\int_0^1\left({1 \over {\ln(1-z)}}+{1 \over z}\right)^4dz=\int_0^1\left(-{1 \over {720}}
+{{53} \over {18}}t-{{57} \over 8}t^2+{{143} \over {36}}t^3+{t^4 \over {24}}\right.$$
$$\left. +\left[{1 \over 6}-{8 \over 3}t+6t^2-{{10} \over 3}t^3\right][\psi(t+1)+\gamma]
\right)dt$$
$$=-{{49} \over {72}} +2 \ln A +{5 \over {2\pi^2}}\zeta(3).  \eqno(2.8)$$

From (2.6) we have
$${1 \over {\ln^5(1-z)}}={1 \over {24}}\sum_{n=1}^\infty \left[(n+1)(n+2)(n+3)(n+4)p_{n+6} 
-2(n+1)(n+2)(n+3)(2n+3)p_{n+5} \right.$$
$$\left. +(n+1)(n+2)(6n^2+12n+7)p_{n+4}-(n+1)(2n+1)(2n^2+2n+1)p_{n+3}+n^4 p_{n+2}\right]z^n$$
$$-{1 \over {24z}}+{5 \over {8z^2}}-{{25} \over {12z^3}}+{5 \over {2z^4}}-{1 \over z^5}, ~~~~~~|z|<1, \eqno(2.9)$$
leading to 
$$\left({1 \over {\ln(1-z)}}+{1 \over z}\right)^5$$
$$={1 \over {24}}\sum_{n=1}^\infty \left[(n+1)(n+2)(n+3)(n+4)p_{n+6} 
-2(n+1)(n+2)(n+3)(2n+3)p_{n+5} \right.$$
$$\left. +(n+1)(n+2)(6n^2+12n+7)p_{n+4}-(n+1)(2n+1)(2n^2+2n+1)p_{n+3}+n^4 p_{n+2}\right]z^n$$
$$+{5 \over 6}\sum_{n=1}^\infty [(n+1)(n+2)(n+3)p_{n+5}-3(n+1)^2(n+2)p_{n+4}+(n+1)(3n^2+3n+1)
p_{n+3}-n^3p_{n+2}]z^{n-1}$$
$$+5\sum_{n=1}^\infty [(n+2)(n+3)p_{n+5}-(n+2)(2n+3)p_{n+4}+(n+1)^2p_{n+3}]z^{n-1}$$
$$+10\sum_{n=1}^\infty [(n+3)p_{n+5}-(n+2)p_{n+4}]z^{n-1}+5\sum_{n=1}^\infty p_{n+5}z^{n-1}.
\eqno(2.10)$$
By performing the integration and using (1.2) we determine
$$\int_0^1\left({1 \over {\ln(1-z)}}+{1 \over z}\right)^5dz
=-{{4367} \over {8640}} +{5 \over 3} \ln A +{{15} \over {8\pi^2}}\zeta(3)-{{35} \over 
3}\zeta'(-3). \eqno(2.11)$$
Here we have used
$$\int_0^1 t^4 \psi(t)dt={{49} \over {180}}-4\zeta'(-1)+\zeta'(0)+6\zeta'(-2)-4\zeta'(-3)$$
$$=-{{11} \over {180}}+4\ln A-{1 \over 2}\ln(2\pi)-{3 \over {2\pi^2}}\zeta(3)-4\zeta'(-3). \eqno(2.12)$$
Based upon the evaluation of $\int_0^1 t^k \psi(t)dt=-k\int_0^1 t^{k-1}\ln \Gamma(t)dt$, $k >-1$, we may anticipate that
$\int_0^1\left({1 \over {\ln(1-z)}}+{1 \over z}\right)^{k+1}dz$ evaluates in terms of
$$\mathbb Q +\mathbb Q\ln A+\mathbb Q{{\zeta(3)} \over \pi^2}+\mathbb Q{{\zeta(5)} \over \pi^4}+\ldots +\mathbb Q {{\zeta(k-1)} \over \pi^{k-2}}$$
$$+\mathbb Q\zeta'(-3)+\mathbb Q \zeta'(-5)+\ldots +\mathbb \zeta'(1-k), ~~~~k ~~\mbox{even},
\eqno(2.13a)$$
and
$$\mathbb Q +\mathbb Q\ln A+\mathbb Q{{\zeta(3)} \over \pi^2}+\mathbb Q{{\zeta(5)} \over \pi^4}+\ldots +\mathbb Q {{\zeta(k)} \over \pi^{k-1}}$$
$$+\mathbb Q\zeta'(-3)+\mathbb Q \zeta'(-5)+\ldots +\mathbb \zeta'(2-k), ~~~~k ~~\mbox{odd}.
\eqno(2.13b)$$

We provide some discussion of intermediate calculations in the above steps.  First, we
consider the integrals $\int_0^1 t^k \psi(t)dt$.  These may be treated by multiple integrations
by parts using the fact that $-\psi(a)$ is the zeroth Stieltjes constant for the Hurwitz zeta
function $\zeta(s,a)$.  That is, we have the limit representation
$$\psi(t)=-\lim_{z\to 1} \left(\zeta(z,t)-{1 \over {z-1}}\right).  \eqno(2.14)$$
Thus,
$$\int_0^1 t^k \psi(t)dt=-\lim_{z\to 1}\int_0^1t^k\left[ \zeta(z,t)-{1 \over {z-1}}\right]dt,  \eqno(2.15)$$
with the interchange justified by uniform convergence of the integral.
We have the properties $\partial_a \zeta(s,a)=-s\zeta(s+1,a)$ and $\int_0^1 \zeta(s,t)dt=0$
for Re $s<1$.  By iteration we may then obtain the integrals
$$\int_0^1 t^k \zeta(z,t)dt=-{1 \over {z-1}}\int_0^1 t^k \partial_t \zeta(z-1,t)dt$$
$$={k \over {z-1}}\int_0^1 t^{k-1}\zeta(z-1,t)dt-{{\zeta(z-1)} \over {z-1}}.  \eqno(2.16)$$

Secondly, the insertion of the integral representation of (1.2) for $p_{n+1}$ into sums
such as (2.10) gives certain hypergeometric summations.  Again, $_pF_q$ denotes the
generalized hypergeometric function.  As an illustration, consider a
contribution from a sum such as $\sum_{n=1}^\infty n^3 p_{n+6}$.  By recalling the property
$(a)_{n+1}=a(a+1)_n$, we have
$$-\sum_{n=1}^\infty n^3{{(-t)_{n+5}} \over {(n+5)!}}=-\sum_{n=0}^\infty (n+1)^3{{(-t)_{n+6}} \over {(n+6)!}}$$
$$=-t(t-1)(t-2)(t-3)(t-4)(t-5)\sum_{n=0}^\infty {{(2)_n^3} \over {(1)_n^3}} {{(6-t)_n} \over
{(n+6)!}}$$
$$=-t(t-1)(t-2)(t-3)(t-4)(t-5){1 \over {720}}\sum_{n=0}^\infty {{(2)_n^3} \over {(1)_n^2}} {{(6-t)_n} \over {(7)_n}}{1 \over {n!}}$$
$$=--t(t-1)(t-2)(t-3)(t-4)(t-5){1 \over {720}} ~_4F_3(2,2,2,6-t;1,1,7;1).  \eqno(2.17)$$

Thirdly, (i) there are relations between the $_{p+1}F_p$ sums so obtained, and (ii) they 
are often related to the digamma function.  By manipulating divided difference forms of the
$\psi$ function, relations such as the following may be obtained:
$$_3F_2(2,2,3-t;3,4;1)+ ~_4F_3(2,2,2,3-t;1,3,4;1)$$
$$=-{{12[3-3t+\gamma t+t\psi(t+1)]} \over {t(t-1)(t-2)}}+{{12[1-2t+\gamma t+t\psi(t+1)]} \over {t(t-1)(t-2)}}={{12} \over {t(t-1)}}.  \eqno(2.18)$$
Likewise, we have
$$_5F_4(2,2,2,2,3-t;1,1,3,4;1)+ ~_4F_3(2,2,2,3-t;1,3,4;1)={{12(4-t)} \over {t(t-1)(t-2)}}.
\eqno(2.19)$$
In this way, summations over the $p_n$ constants may be evaluated.

We also record the following.
{\newline \bf Lemma 3}. For Re $x>0$,
$$\psi(x)-\ln x=-\sum_{n=1}^\infty p_{n+1} {{(n-1)!} \over {(x)_n}}.  \eqno(2.20)$$
{\bf Corollary 2}.  For Re $x>-1$,
$$x(\psi(x)-\ln x)=-\sum_{n=0}^\infty p_{n+2} {{n!} \over {(x+1)_n}}.  \eqno(2.21)$$

{\it Proof}.  For Re $x>0$,
$$\psi(x)-\ln x=\int_0^\infty e^{-xt}\left({1 \over t}-{1 \over {1-e^{-t}}}\right)dt.
\eqno(2.22)$$
Let $t=-\ln(1-z)$, giving
$$\psi(x)-\ln x=-\int_0^1 (1-z)^{x-1} \left({1 \over {\ln(1-z)}}+{1 \over z} \right)dz.  \eqno(2.23)$$
Now use the generating function (1.3), so that
$$\psi(x)-\ln x=-\sum_{n=1}^\infty p_{n+1}\int_0^1 (1-z)^{x-1} z^{n-1} dz$$
$$=-\sum_{n=1}^\infty p_{n+1} B(x,n).  \eqno(2.24)$$

For Corollary 2, we use $x/(x)_n=1/(x+1)_{n-1}$.  \qed

The proof we have given of (2.20) complements that of Proposition 5(a) of \cite{coffeyjnt}.
Of course this relation may be directly verified with the aid of (1.3):
$$-\sum_{n=1}^\infty p_{n+1} {{(n-1)!} \over {(x)_n}}=\int_0^1 \sum_{n=1}^\infty 
{{(-t)_n} \over {n(x)_n}}dt$$
$$=\int_0^1[\psi(x)-\psi(t+x)]dt=\psi(x)-\ln[\Gamma(x+1)/\Gamma(x)]=\psi(x)-\ln x.  \eqno(2.25)$$

We may note that the representation (2.23) may be repeatedly integrated by parts.  We have for example
$$\psi(x)-\ln x=-{1 \over x}\int_0^1 (1-z)^{x-1}\left[{1 \over {\ln^2(1-z)}}-{1 \over z^2}+
{1 \over z}\right]dz-{1 \over {2x}}$$
$$=-{1 \over x^2}\int_0^1 (1-z)^{x-1}\left[{2 \over {\ln^3(1-z)}}+{2 \over z^3}-{3 \over z^2}+
{1 \over z}\right]dz-{1 \over {2x}}-{1 \over {12x^2}}$$
$$=-{1 \over x^3}\int_0^1 (1-z)^{x-1}\left[{6 \over {\ln^4(1-z)}}-{6 \over z^4}+{{12} \over z^3}-{7 \over z^2}+{1 \over z}\right]dz-{1 \over {2x}}-{1 \over {12x^2}}$$
$$=-{1 \over x^4}\int_0^1 (1-z)^{x-1}\left[{{24} \over {\ln^5(1-z)}}+{{24} \over z^5}-{{60} \over z^4}+{{57} \over z^3}-{{22} \over z^2}+{1 \over z}\right]dz-{1 \over {2x}}-{1 \over {12x^2}}+{1 \over {120x^4}}.  \eqno(2.26)$$
In writing these equations, we have used the values of $p_2$, $p_3$, $p_4$, and $p_5$ for the boundary terms.  We note that the latter terms give the asymptotic expansion of $\psi(x)-\ln x$
as $x \to \infty$.  We expand on this point next with regard to Corollary 3.

The asymptotic expansion 
$$\psi(z)-\ln z=-{1 \over {2z}}-\sum_{n=1}^\infty {B_{2n} \over {2n z^{2n}}}
=-{1 \over {2z}}-{1 \over {12z^2}}+{1 \over {120z^4}}-{1 \over {252 z^6}} + \ldots, \eqno(2.27)$$
where $B_k$ are the Bernoulli numbers, is well known.  In fact, we may readily derive it in
the following manner.  We have
$$\tilde{\psi}(x) \equiv \psi(x)-\ln x+{1 \over {2x}}=-\int_0^\infty e^{-2tx}\left(\coth t -{1
\over t}\right)dt. \eqno(2.28)$$
The asymptotic form of the integral for large $x$ is obtained as $t \to 0$, in which case we 
may use the expansion
$$\coth t -{1 \over t} = \sum_{k=1}^\infty {2^{2k} \over {(2k)!}}B_{2k}t^{2k-1}, ~~~~t^2 < 
\pi^2.  \eqno(2.29)$$
Then (2.27) follows.  
Now we may apply Stirling's formula for $\Gamma$ to Lemma 2, so that for $x \to \infty$,
$$\psi(x)-\ln x =-\sum_{n=1}^\infty p_{n+1}{{(n-1)!} \over x^n}\left[1+{{n(1-n)} \over {2x}}
+{{n(2-3n-2n^2+3n^3)} \over {24x^2}}\right.$$
$$\left.+{{n^2(-2+n+3n^2-n^3-n^4)} \over {48x^3}}+ \ldots \right]. \eqno(2.30)$$
Therefore, by matching asymptotic expansions, we have the following.
{\newline \bf Corollary 3}.  (a) $p_3={1 \over {12}}={B_2 \over 2}$ and
${B_4 \over 4}=6p_5+p_3-6p_4$.  (b) ${B_{2n} \over {2n}}$ may be expressed as a sum
of $p_n$ values with rational coefficients. 

Elaborating part (b), we have the following.  We let $S(n,k)$ denote the Stirling numbers of
the second kind.
{\newline \bf Proposition 1}.
$${B_n \over n}=\sum_{k=1}^n (-1)^k (k-1)!S(n,k)\sum_{\ell=0}^{k-1} p_{\ell+2}$$
$$=\sum_{\ell=0}^{n-1} p_{\ell+2}\sum_{k=\ell+1}^n (-1)^k(k-1)!S(n,k). \eqno(2.31)$$

{\it Proof}.  We let $B_k^{(k)}=(-1)^k \int_0^1 (x)_kdx$ be the N\"{o}rlund numbers
(e.g., \cite{norlund,coffeyseries}), such that $B_0^{(0)}=1$, $B_1^{(1)}=-1/2$, and $B_2^{(2)}=5/6$, and have
$$B_n^{(n)}+nB_{n-1}^{(n-1)}=(-1)^{n+1}n!p_{n+1}. \eqno(2.32)$$
By iterating, using the initial value $B_0^{(0)}=1$, we obtain
$$B_n^{(n)}=(-1)^n n!\left(1-\sum_{k=0}^{n-1} p_{k+2}\right). \eqno(2.33)$$
Substituting into
$$\sum_{k=1}^n S(n,k){B_k^{(k)} \over k}=-{B_n \over n}, \eqno(2.34)$$
and using the sum
$$\sum_{k=1}^n (-1)^k (k-1)!S(n,k)=0 \eqno(2.35)$$
gives the Proposition. \qed

The integral representation (2.23) also leads to the following series representation.
{\newline \bf Proposition 2}.  For Re $y>0$,
$$\ln \Gamma(y)-y\ln y+y=-{1 \over {6y}}-{1 \over 2}\psi(y)-\sum_{n=1}^\infty[(n+2)p_{n+3}
-np_{n+2}]B(y,n+1) + {1 \over 2}\ln(2\pi).  \eqno(2.36)$$

{\it Proof}.  We first integrate (2.23) from $x=1$ to $y$, giving
$$\ln\Gamma(y)-y\ln y+y-1=-\int_0^1{{[(1-z)^{y-1}-1]} \over {\ln(1-z)}}\left({1 \over {\ln(1-z)}}-{1 \over z}\right)dz$$
$$=-\int_0^1 [(1-z)^{y-1}-1]\left\{\sum_{n=1}^\infty[(n+1)p_{n+3}-np_{n+2}]z^n+p_3-{1 \over z}+\sum_{n=1}^\infty p_{n+1}z^{n-2}\right\}dz$$
$$=-\int_0^1 [(1-z)^{y-1}-1]\left\{\sum_{n=1}^\infty[(n+1)p_{n+3}-np_{n+2}]z^n+p_3
-{1 \over {2z}} +\sum_{n=1}^\infty p_{n+2}z^{n-1}\right\}dz, \eqno(2.37)$$
using (1.3) and (1.4).  Performing the integration, we find
$$\ln\Gamma(y)-y\ln y+y-1=-\sum_{n=1}^\infty[(n+1)p_{n+3}-np_{n+2}]\left[B(y,n+1)-{1 \over {n+1}}\right]-p_3\left({1 \over y}-1\right)$$
$$-{1 \over 2}[\psi(y)+\gamma]
  -\sum_{n=1}^\infty p_{n+2}\left[B(y,n)-{1 \over n}\right].  \eqno(2.38)$$
Next, the integral representation (1.2) is used for the sums absent the Beta function,
$$\sum_{n=1}^\infty {p_{n+2} \over n}={1 \over 2}\ln(2\pi)-1-\gamma,  \eqno(2.39)$$
$$\sum_{n=1}^\infty p_{n+3}={5 \over {12}}, ~~~~ -\sum_{n=1}^\infty {n \over {n+1}}p_{n+2}
=\gamma-1.  \eqno(2.39b)$$
Alternatively, these sums may be obtained by integrating and other otherwise manipulating
the generating function (1.3) and taking $z \to 1$.  Then combining terms of (2.38) and (2.39) gives the Proposition.  \qed

The coefficients $p_{n+1}$ may be readily related to other quantities, including the
Bernoulli numbers of the second kind $b_n$ ($n \geq 0$) \cite{jordan,roman},
$$b_n=\int_0^1 {{\Gamma(t+1)} \over {\Gamma(t-n+1)}}dt,  \eqno(2.40)$$
with $b_0=1$, $b_1=1/2$, $b_2=-1/6$, and $b_3=1/4$.  We have 
$$p_{n+1}=-{1 \over {n!}}\int_0^1 {{\Gamma(n-t)} \over {\Gamma(-t)}}dt
={{(-1)^{n+1}} \over {n!}}\int_0^1 {{\Gamma(t+1)} \over {\Gamma(t+1-n)}}dt, \eqno(2.41)$$
and hence $p_{n+1}=(-1)^{n-1}b_n/n!$.

The following gives a series representation for $\ln A$.  The method of proof shows that
a family of such series may be written.
{\newline \bf Proposition 3}.  
$$\ln A={1 \over 4}+\sum_{n=1}^\infty \left\{{1 \over 2}(n+2)p_{n+4}+(n+1)\left({1 \over n}-1\right) p_{n+3}+\left[{n^2 \over {2(n+1)}}-1+{{(6n+1)} \over {12(n+1)}}\right]p_{n+2}\right\}.  \eqno(2.42)$$

{\it Proof}.  Let 
$\zeta(s,a)$ be the Hurwitz zeta function.  Then for Re $s>-(2n-1)$, $n \in \mathbb N_0$, and Re $a>0$, there is the integral 
representation
$$\zeta(s,a)=a^{-s}+\sum_{k=0}^n (s)_{k-1}{B_k \over {k!}}a^{-k-s+1}
+{1 \over {\Gamma(s)}}\int_0^\infty \left({1 \over {e^t-1}}-\sum_{k=0}^n {B_k \over {k!}}
t^{k-1}\right)e^{-at}t^{s-1}dt.  \eqno(2.43)$$
By taking $n=2$ and $a=1$, one may find
$$\ln A={1 \over 4}+\int_0^\infty\left({1 \over {e^t-1}}-{1 \over t}+{1 \over 2}-{t \over {12}}
\right){e^{-t} \over t^2}dt.  \eqno(2.44)$$
Changing variable with $t=-\ln(1-z)$ we have
$$\ln A={1 \over 4}+\int_0^1\left({1 \over z}+{1 \over {\ln(1-z)}}-{1 \over 2}+{{\ln(1-z)} \over
{12}}\right){{dz} \over {\ln^2(1-z)}}.  \eqno(2.45)$$
Next the generating functions (1.3), (1.4), and (2.1) are employed in the integrand, and all
terms $O(z^{-k})$, $k=1,2,3$ are nullified, as they should.  Performing the integration gives
$$\ln A={1 \over 4}+\sum_{n=1}^\infty \left[\left({{n+1} \over n}\right)p_{n+3}-p_{n+2}\right]$$
$$+{1 \over 2}\sum_{n=1}^\infty\left[(n+2)p_{n+4}-(2n+1)p_{n+3}+{n^2 \over {n+1}}p_{n+2}\right]$$
$$-{1 \over 2}\sum_{n=1}^\infty\left[p_{n+3}-{n \over {n+1}}p_{n+2}+p_3\right]+{1 \over {12}}
\sum_{n=1}^\infty {p_{n+1} \over n}.  \eqno(2.46)$$
Shifting the index on the last sum and combining terms gives the Proposition. \qed

Noting that $\ln A \simeq 0.248754477033784262547253$, the summation in (2.29) provides an
appropriate correction to $1/4$.

{\bf Corollary 4}.  The constant 
$$\zeta'(2)=\zeta(2)[\gamma+\ln(2\pi)-12\ln A] \eqno(2.47)$$
may be written in terms of the coefficients $p_n$.

{\it Proof}.  The relation of $\zeta'(2)$ to $\zeta'(-1)$ follows from the functional
equation of the Riemann zeta function.  The value $\zeta(2)$ may be found in terms of 
$p_n$'s via the integral representations (2.34) at $a=1$.  The constant $\ln(2\pi)$ may be
written in terms of $p_n$'s via (1.7) and (1.9).  Finally, the Euler constant
$\gamma=\sum_{n=1}^\infty {p_{n+1} \over n}$, as $\gamma=\int_0^1\left({1 \over {\ln x}}+
+{1 \over {1-x}}\right)dx$.  
    
We may note that more generally we may similarly write series representations for the 
logarithm of the double Gamma function $\Gamma_2$, since we have for Re $a>0$
$$\ln \Gamma_2(a)=\ln A-{a^2 \over 4}+\left({a^2 \over 2}-{a \over 2}+{1 \over {12}}\right)
\ln a+(1-a)\ln\Gamma(a)$$
$$-\int_0^\infty\left({1 \over {e^t-1}}-{1 \over t}+{1 \over 2}-{t \over {12}}\right){e^{-at}
\over t^2}dt$$
$$=\ln A-{a^2 \over 4}+\left({a^2 \over 2}-{a \over 2}+{1 \over {12}}\right)
\ln a+(1-a)\ln\Gamma(a)$$
$$-\int_0^1\left({1 \over z}+{1 \over {\ln(1-z)}}-{1 \over 2}+{{\ln(1-z)} \over
{12}}\right)(1-z)^{a-1}{{dz} \over {\ln^2(1-z)}}.  \eqno(2.48)$$
We arrive at 
{\newline \bf Proposition 4}.  For Re $a>0$,
$$\ln \Gamma_2(a)=\ln A-{a^2 \over 4}+\left({a^2 \over 2}-{a \over 2}+{1 \over {12}}\right)
\ln a+(1-a)\ln\Gamma(a)$$
$$=-\sum_{n=1}^\infty \left[(n+2)^2p_{n+4}-(n+1)(n+2)p_{n+3}+{1 \over 2}\left(n^2+n+{1 \over 6}\right)p_{n+2}\right]B(n+1,a).  \eqno(2.49)$$
The details of this calculation are omitted.

We conjecture that the following inequalities hold for the coefficients $p_n$.
{\newline \bf Conjecture 1}.  (i) For $n \geq 2$,
$$(n+1)p_{n+3}-np_{n+2} < 0, \eqno(2.50)$$
(ii) that the sequence $\{p_n\}$ is strictly convex, i.e., for $n \geq 3$,
$$p_n < {1 \over 2}(p_{n+1}+p_{n-1}), \eqno(2.51)$$
(iii) that the sequence $\{p_n\}$ is strictly log-convex, i.e., for $n \geq 3$,
$$p_n^2 < p_{n-1}p_{n+1}.  \eqno(2.52)$$
The convexity and log-convexity themselves are probably not difficult to show, and (iii) implies (ii). 
Certainly these inequalities hold for $n$ sufficiently large.  Using the
known asymptotic form $p_n \sim 1/[n(\ln n+ \gamma)^2]$, the leading asymptotic form
of the difference of the left and right sides of the Conjecture is given by:  for (i), $-2/[n(\ln n+\gamma)^3]$, for (ii), $-\ln^2 n/[n^3(\ln n+\gamma)^4]$, and for (iii), 
$-\ln^2 n/[n^4(\ln n+\gamma)^6]$.

\medskip
\centerline{\bf Expressions for the Stieltjes constants and $\zeta(m,a)$ values}
\medskip

The Hurwitz zeta function, defined by $\zeta(s,a)=\sum_{n=0}^\infty (n+a)^{-s}$
for Re $s>1$ and Re $a>0$ extends to a meromorphic function in the entire
complex $s$-plane.  In the Laurent expansion 
$$\zeta(s,a)={1 \over {s-1}}+ \sum_{n=0}^\infty {{(-1)^n} \over {n!}}\gamma_n(a)
(s-1)^n, \eqno(3.1)$$
$\gamma_n(a)$ are the Stieltjes constants (e.g., \cite{coffey2009,coffeyjnt}), 
and by convention one takes $\gamma_k = \gamma_k(1)$.  One has $\gamma_0(a)=-\psi(a)$
and $\gamma_0=\gamma$.  The Stieltjes constants may be expressed via the limit formula
$$\gamma_k(a)=\lim_{N \to \infty}\left[\sum_{n=0}^N {{\ln^k(n+a)} \over {n+a}}- {{\ln^{k+1}
(N+a)} \over {k+1}}\right].$$
We dispense with further preliminaries concerning the $\gamma_n(a)$'s.

We may write series and integral representations for these constants based upon the 
$p_n$ coefficients.
This development is illustrated in the next result.
{\newline \bf Proposition 5}. (a)
$$-{1 \over 2}\ln^2 a-\gamma_1(a)=\gamma[\ln a-\psi(a)]+\sum_{n=1}^\infty p_{n+1}
\int_0^1 u^{a-1} \ln(-\ln u)(1-u)^{n-1}du$$
$$=\gamma[\ln a-\psi(a)]+\int_0^1 u^{a-1}\ln(-\ln u)\left[{1 \over {1-u}}+{1 \over {\ln u}}
\right]du, \eqno(3.2a)$$
and
$$-\gamma_1-\gamma^2=\sum_{n=1}^\infty p_{n+1}\int_0^\infty e^{-t} \ln t(1-e^{-t})^{n-1}dt$$
$$=-\int_0^\infty \left[{1 \over {1-e^t}}+{e^{-t} \over t}\right]\ln t ~dt,  \eqno(3.2b)$$
(b)
$$\gamma_2=-\gamma(\gamma^2+\zeta(2)+2\gamma_1)+\int_0^\infty (\ln^2 t) e^{-t} \sum_{n=1}^\infty
p_{n+1}(1-e^{-t})^{n-1} dt, \eqno(3.3)$$
and (c)
$$-\gamma_3=\gamma^4+{\pi^2 \over 2}\gamma_1+{\gamma^2 \over 2}(\pi^2+6\gamma_1)+3\gamma \gamma_2+2\gamma\zeta(3)$$
$$+\int_0^\infty (\ln^3 t) e^{-t} \sum_{n=1}^\infty p_{n+1}(1-e^{-t})^{n-1}dt.  \eqno(3.4)$$

{\it Proof}.  For Re $s>0$ and Re $a>0$ we have the integral representation
$$\zeta(s,a)-{1 \over {(s-1)}}{1 \over a^{s-1}}={1 \over {\Gamma(s)}}\int_0^\infty e^{-ax}
x^{s-1}\left({1 \over {1-e^{-x}}}-{1 \over x}\right)dx$$
$$={1 \over {\Gamma(s)}}\int_0^1 (1-z)^{a-1}[-\ln(1-z)]^{s-1}\left[{1 \over z}+{1 \over {\ln(
1-z)}}\right]dz$$
$$={1 \over {\Gamma(s)}}\sum_{n=1}^\infty p_{n+1}\int_0^1 u^{a-1}(-\ln u)^{s-1} (1-u)^{n-1}du.
\eqno(3.5)$$

It is readily found that the limit of (3.5) as $s \to 1$ agrees with Lemma 2, and this
result is used in all parts of the Proposition.  For parts (a)-(c) we take successive
derivatives of (3.5) with respect to $s$ and evaluate at $s=1$.  For the integral
representations in (a) we use the integral representation for $p_{n+1}$ in (1.2). \qed

The following further emphasizes the connection of the $p_{n+1}$ coefficients with analytic
number theory.  We let $\psi'$ be the trigamma function, $\psi^{(j)}$ the polygamma functions, and $H_n=\sum_{k=1}^n {1 \over k}=\psi(n+1)+\gamma$ be the $n$th harmonic number.
{\newline \bf Corollary 5}.
$$\zeta(2)=1+\sum_{n=1}^\infty {p_{n+1} \over n}H_n, \eqno(3.6)$$
and
$$\zeta(3)={1 \over 2}+{{\zeta(2)} \over 2}\gamma+{1 \over 2}\sum_{n=1}^\infty {p_{n+1} \over
n}[H_n^2-\psi'(n+1)].  \eqno(3.7)$$

{\it Proof}.  This follows from the special case of (3.5),
$$\zeta(s)-{1 \over {(s-1)}}={1 \over {\Gamma(s)}}\sum_{n=1}^\infty p_{n+1}\int_0^1 (-\ln u)^{s-1} (1-u)^{n-1}du.  \eqno(3.8)$$
The integrals are given by
$$\int_0^1 (-\ln u)^{s-1} (1-u)^{n-1}du=(-1)^{s-1} \partial_x^{s-1}\left.B(x,n)\right|_{x=1}.
\eqno(3.9)$$
In writing (3.7) we have used the previously given summation expession for $\gamma$. \qed

Similarly further values of $\zeta(n)$ may be written.  We may note the appearance of
generalized harmonic numbers $H_n^{(r)}=\sum_{k=1}^n {1 \over k^r}$, $H_n \equiv H_n^{(1)}$.
In particular, as regards (3.7), $H_n^{(2)}=-[\psi'(n+1)-\psi'(1)]=-[\psi'(n+1)-\zeta(2)]$,
so that
$$\zeta(3)={1 \over 2}+{1 \over 2}\sum_{n=1}^\infty {p_{n+1} \over n}[H_n^2+H_n^{(2)}].
\eqno(3.10)$$
Furthermore,
$$\zeta(4)={1 \over 3}+{1 \over 6}\sum_{n=1}^\infty {p_{n+1} \over n}[H_n^3+3H_nH_n^{(2)}+
2H_n^{(3)}].  \eqno(3.11)$$
As in Corollary 5, the summand is positive and the convergence is from below.

The generalized harmonic numbers are given by
$$H_n^{(r)}={{(-1)^{r-1}} \over {(r-1)!}}\left[\psi^{(r-1)}(n+1)-\psi^{(r-1)}
(1) \right ]. \eqno(3.12)$$
We may systematize Corollary 5 and related results in the following manner.  We introduce
the (exponential) complete Bell polynomials $Y_n=Y_n(x_1,x_2,\ldots,x_n)$ appearing in the
expansion
$$\exp\left(\sum_{m=1}^\infty x_m {t^m \over {m!}}\right)=1+\sum_{n=1}^\infty Y_n(x_1,x_2,\ldots,x_n){t^n \over {n!}}, ~~~~Y_0=1.  \eqno(3.13)$$
Then
{\newline \bf Proposition 6}.  For $m \in \mathbb{N}$, $m \geq 2$, (a)
$$\zeta(m)={1 \over {m-1}}+{1 \over {(m-1)!}}\sum_{n=1}^\infty {p_{n+1} \over n} Y_{m-1}
(H_n,H_n^{(2)},2!H_n^{(3)},\ldots,(m-2)!H_n^{(m-1)}),  \eqno(3.14)$$
and (b) for Re $a>0$,
$$\zeta(m,a)={1 \over {(m-1)}}{1 \over a^{m-1}}+{{(-1)^{m-1}} \over {(m-1)!}}\sum_{n=1}^\infty p_{n+1}B(a,n)Y_{m-1}[\psi(a)-\psi(a+n),\psi'(a)-\psi(a+n),\ldots,$$
$$\psi^{(m-1)}(a)-\psi^{(m-1)}(a+n)].  \eqno(3.15)$$

{\it Proof}.  We apply (e.g., \cite{coffeyutilitas})
{\newline \bf Lemma 4}.  For differentiable functions $f$ and $g$ such that
$f'(x)=f(x)g(x)$, assuming all higher order derivatives exist, we have
$$\left({d \over {dx}}\right)^j f(x) = f(x) Y_j\left[g(x),g'(x),\ldots,g^{(j-1)}
(x)\right].  \eqno(3.16)$$
We put $f(x)=B(x,n)$ and $g(x)={d \over {dx}}\ln B(x,n)=\psi(x)-\psi(x+n)$.  Then by (3.12)
$g^{(r)}(1)=(-1)^r (r-1)!H_n^{(r)}$.  Part (a) then follows from (3.8) and (3.9) with $s=m$ 
and $B(1,n)=1/n$.

Similarly for part (b) we use (3.5), in which the integral on the right side is given by
$(-1)^{s-1}\partial_a^{s-1}B(a,n)$. \qed

Alternatively, the integral on the right side of (3.5) may be treated with a generating
function for the Stirling numbers of the first kind \cite{nbs} (p. 824) so that
$$\int_0^1 (1-u)^{a-1}[-\ln(1-u)]^{m-1}u^{n-1}du
=(-1)^{m-1}(m-1)!\sum_{j=m-1}^\infty {{(-1)^j} \over {j!}}s(j,m-1)B(n+j,a).  \eqno(3.17)$$

\noindent
{\it Additional sums}

We collect the following summations with the $p_{n+1}$ coefficients.
{\newline \bf Proposition 7}.
$$\sum_{n=1}^\infty {p_{n+1} \over {n+a}}={1 \over a}-\int_0^1 B(a,x+1)dx, \eqno(3.18)$$
in particular
$$\sum_{n=1}^\infty {p_{n+1} \over {n+1}}=1-\ln 2, \eqno(3.19)$$
$$\sum_{n=1}^\infty {p_{n+1} \over n^2}={1 \over 2}(\gamma^2-1)+{\pi^2 \over {12}}+{1 \over 2}
\int_0^1 \psi^2(x+1)dx, \eqno(3.20)$$
$$\sum_{n=1}^\infty {p_{n+1} \over n^3}={1 \over {12}}[-5+2\gamma^3+\gamma \pi^2+4\zeta(3)]
+{1 \over 6}\int_0^1 [3\gamma \psi^2(x+1)+\psi^3(x+1)]dx,  \eqno(3.21)$$
and for $j \in \mathbb{N}^+$,
$$\sum_{n=1}^\infty {p_{n+1} \over n^j}z^n=z \int_0^1 x ~_{j+2}F_{j+1}(1,1,\ldots,1,1-x;2,2,
\ldots,2;z)dx.  \eqno(3.22)$$

{\it Proof}.  These may be obtained with the aid of the integral representation of (1.2), and
we omit further details. \qed

As concerns the integral on the right side of (3.20), we have the following result.
{\bf Corollary 6}.  
$$\int_0^1 \psi^2(x+1)dx=2\gamma_1-{\pi^2 \over 3}+\int_0^1\left[2{{\psi(x)} \over x}+{1 \over
x^2}-2\gamma \psi(x)+\psi'(x)\right]dx.  \eqno(3.23)$$

{\it Proof}.  We use Proposition 3(b) of \cite{coffeyseries},
$$\gamma_1={\pi^2 \over 6}+\int_0^1 \left(\gamma \psi(x)+{1 \over 2}[\psi^2(x)-\psi'(x)]
\right)dx, \eqno(3.24)$$  
together with $\psi^2(x+1)=\psi^2(x)+2{{\psi(x)} \over x}+{1 \over x^2}$.  \qed

Based upon various representations of the Beta function, we may collect the following
representations for the integral occurring in (3.5) for the Riemann zeta function case,
$a=1$.  It should be clear which of these are restricted to $s \geq 1$ an integer.
{\newline \bf Proposition 8}.  
$$I_n(s) \equiv \int_0^1 (-\ln u)^{s-1} (1-u)^{n-1}du=\left. (-1)^{s-1}\partial_a^{s-1}
B(a,n)\right|_{a=1}$$
$$={{\Gamma(s)} \over n}\sum_{k=1}^n {{(-1)^{k-1}} \over k^{s-1}}{n \choose k}$$
$$={{\Gamma(s)} \over n}\sum_{1\leq j_1\leq \cdots \leq j_{s-1}\leq n}{1 \over {j_1 \cdots j_{s-1}}}$$
$$=\Gamma(s)\int_{[0,1]^s} (1-x_1x_2\cdots x_s)^{n-1} dx_1\cdots dx_s$$
$$={1 \over {2\pi i}}{{\Gamma(s)} \over n}\oint_{|z|=r<1} {1 \over z^s} \prod_{j=1}^n {{dz} \over {(1-z/j)}}$$
$$={{(-1)^n} \over {2\pi i}}{{\Gamma(s)} \over n} \int_{1/2-i\infty}^{1/2+i\infty} {{n!} \over
{y^s (y-1)\cdots (y-n)}}dy.  \eqno(3.25)$$
Furthermore, for $s \geq 2$,
$${{(s-1)} \over n}I_n(s)=\left({{s-1} \over {n-1}}\right)I_{n-1}(s)+{1 \over n}I_n(s-1),
\eqno(3.26a)$$
and
$${{(s-1)} \over n}I_n(s)=\sum_{j=1}^n {{I_j(s-1)} \over j^2}.  \eqno(3.26b)$$

\medskip
\centerline{\bf Upper bound for $p_{n+1}$ and other results, including}
\centerline{\bf polylogarithmic representation of the Stieltjes constants}
\medskip

We let Si$(z)=\int_0^z {{\sin t} \over t}dt$ be the sine integral.
{\newline \bf Proposition 9}. For $n \geq 1$,
$$p_{n+1}< -{1 \over 2}+{1 \over \pi}\mbox{Si}(\pi)+(-1)^{n-1}\left[{1 \over 2}-{{\mbox{Si}
[(n-1)\pi]} \over \pi}\right].  \eqno(4.1)$$

{\it Proof}.  Based upon contour integration, Knessl has shown the following integral
representation \cite{coffey2009},
$$p_{n+1}=\int_0^\infty {1 \over {(1+u)^n}} {{du} \over {(\ln^2 u+\pi^2)}},
~~~~~~ n \geq 1.  \eqno(4.2)$$
Then
$$p_{n+1}=\int_0^1 {1 \over {(1+u)^n}} {{du} \over {(\ln^2 u+\pi^2)}}+\int_1^\infty {1 \over {(1+u)^n}} {{du} \over {(\ln^2 u+\pi^2)}}$$
$$<\int_0^1 {{du} \over {(\ln^2u+\pi^2)}}+\int_1^\infty {{du} \over {u^n(\ln^2u+\pi^2)}}$$
$$=\int_1^\infty \left({1 \over u^2}+{1 \over u^n}\right){{du} \over {(\ln^2u+\pi^2)}}$$
$$=\int_0^\infty {{[1+e^{-(n-1)v}]} \over {v^2+\pi^2}}dv \eqno(4.3)$$
$$=-{1 \over 2}+{1 \over \pi}\mbox{Si}(\pi)+(-1)^{n-1}\left[{1 \over 2}-{{\mbox{Si}
[(n-1)\pi]} \over \pi}\right].$$
The integral of (4.3) could be evaluated by means of contour integration, but we supply
another means.  We put
$$I(k) \equiv \int_0^\infty {e^{-(k-1)v} \over {v^2+\pi^2}}dv, \eqno(4.4)$$
and form
$$I''(k)+\pi^2 I(k)=\int_0^\infty e^{-(k-1)v}dv={1 \over {k-1}}.  \eqno(4.5)$$
The homogeneous solutions of this differential equation are of course $\cos k\pi$ and
$\sin k\pi$ and their constant Wronskian is $W=\pi$.  Per variation of parameters, a particular solution then takes the form
$$I_p(k)=-{{\cos \pi k} \over \pi}\int {{\sin \pi k} \over {k-1}}dk+{{\sin \pi k} \over \pi}
\int {{\cos \pi k} \over {k-1}}dk.$$
Then solving (4.5) subject to $I(1)=1/2$ gives
$$I(k)={1 \over \pi}\left\{-\sin(k\pi)\mbox{Ci}[(k-1)\pi]+\cos(k\pi)\mbox{Si}[(k-1)\pi]
\right\}-{1 \over 2}\cos k\pi + c_2\sin k\pi, \eqno(4.6)$$
where Ci$(z)=-\int_z^\infty {{\cos t} \over t}dt$ is the cosine integral.  Then imposing
$I(\infty)=0$ gives the constant $c_2=0$. \qed

Asymptotically, as $n \to \infty$ on the right side of (4.1), this upper bound is
${{\mbox{Si}(\pi)}\over \pi}-{1 \over 2}+{1 \over {\pi^2 n}}+O\left({1 \over n^2}\right)$.

We may extend Lemma 3 to the following.
{\newline \bf Lemma 5}.
$$\psi(x)-\ln x=-\sum_{n=1}^\infty p_{n+1}{{(n-1)!} \over {(x)_n}}=-{1 \over {2x}}-2
\int_0^\infty {{t dt} \over {(t^2+x^2)(e^{2\pi t}-1)}}$$
$$=-{1 \over {2x}}-2\int_0^\infty {{vdv} \over {(1+v^2)(e^{2\pi x v}-1)}}$$
$$-{1 \over {2x}}-2\int_1^\infty {{\ln u ~du} \over {u(1+\ln^2u)(u^{2\pi x}-1)}}$$
$$=-{1 \over x}\int_0^\infty {{~_2F_1(1,1;x+1;v)dv} \over {v\left[\ln^2\left({1 \over v}-1
\right)+\pi^2\right]}}.  \eqno(4.7)$$

{\it Proof}.  The first line follows from a known integral representation of $\psi(x)-\ln x$
\cite{grad} (p. 943).  The final equality follows from the representation (4.2) for $p_{n+1}$.
\qed 

For the next result we introduce the polylogarithm function Li$_s(z)=\sum_{k=1}^\infty {z^k 
\over k^s}$ that may be analytically continued to the whole complex plane.  This function has
the integral representation for Re $s>0$
$$\mbox{Li}_s(z)={z \over {\Gamma(s)}}\int_0^\infty {{t^{s-1} dt} \over {e^t -z}}, \eqno(4.8)$$
and we note that Li$_1(z)=-\ln(1-z)$. 
The branch cut for Li$_s(z)$ in the complex $z$ plane may be taken from $1$ to $\infty$.  
This function is also given when $s$ is a positive integer by Li$_m(z)=z ~_{m+1}F_m(1,1,\ldots,
1;2,\ldots,2;z)$, and this provides another way of seeing the branch point at $z=1$.

{\bf Proposition 10}.  
$$\gamma_1+\gamma^2=\int_0^\infty \left[\gamma \ln\left(1+{1 \over u}\right)+\left. \partial_s \right|_{s=1}\mbox{Li}_s\left(-{1 \over u}\right)\right]{{du} \over {(\ln^2 u+\pi^2)}},  \eqno(4.9)$$
and
$$\gamma_2+\gamma[\gamma^2+\zeta(2)+2\gamma_1]$$
$$=\int_0^\infty \left[(\gamma^2+\zeta(2)) \ln\left(1+{1 \over u}\right)+2\gamma\left. \partial_s \right|_{s=1}
\mbox{Li}_s\left(-{1 \over u}\right)-\left. \partial_s^2 \right|_{s=1}
\mbox{Li}_s\left(-{1 \over u}\right)\right]{{du} \over {(\ln^2 u+\pi^2)}}.  \eqno(4.10)$$

{\it Proof}.  By (3.2b) and the representation (4.2),
$$-\gamma_1-\gamma^2=\sum_{n=1}^\infty \int_0^\infty {1 \over {(1+u)^n}}{{du} \over {(\ln^2 u+
\pi^2)}}\int_0^\infty e^{-t} \ln t(1-e^{-t})^{n-1}dt$$
$$=\int_0^\infty \int_0^\infty {{du} \over {(1+e^t u)}} \ln t ~dt.$$
The interchanges are justified by the absolute convergence of the integrals.
Then (4.9) follows from logarithmic differentiation of the integral of (4.8). 
Similarly for (4.10), (3.3) is used along with (4.2).  \qed

Several versions of (4.10) may be written by employing (4.9).

Proposition 10 is clarified and generalized with the following.
{\newline \bf Proposition 11}.  We have
$$\zeta(s)-{1 \over {s-1}}=-\int_0^\infty {{\mbox{Li}_s(-v)dv} \over {v^2(\ln^2 v+\pi^2)}}.
\eqno(4.11)$$
Consequently, for $k \geq 0$,
$$\gamma_k=(-1)^{k-1} \left.\left({\partial \over {\partial s}}\right)^k \right|_{s=1}
\int_0^\infty {{\mbox{Li}_s(-v)dv} \over {v^2(\ln^2 v+\pi^2)}}.  \eqno(4.12)$$

{\it Proof}.  We first combine the representation (3.5) with $a=1$ with the second line
of Proposition 8 resulting from binomial expansion:
$$\zeta(s)-{1 \over {s-1}}=\sum_{n=1}^\infty {p_{n+1} \over n}\sum_{k=1}^n {n \choose k}
{{(-1)^{k-1}} \over k^{s-1}}$$
$$=\sum_{n=0}^\infty {p_{n+2} \over {(n+1)}}\sum_{k=0}^n {{n+1} \choose {k+1}} {{(-1)^k} \over
{(k+1)^{s-1}}}=\sum_{n=0}^\infty p_{n+2}\sum_{k=0}^n {n \choose k} {{(-1)^k} \over {(k+1)^s}}$$
$$=\sum_{k=0}^\infty \sum_{n=k}^\infty p_{n+2} {n \choose k} {{(-1)^k} \over {(k+1)^s}}.  \eqno(4.13)$$
We now use the representation (4.2) for $p_{n+2}$, so that
$$\zeta(s)-{1 \over {s-1}}=\sum_{k=0}^\infty {{(-1)^k} \over u^{k+1}}{1 \over {(k+1)^s}}
{{du} \over {(\ln^2 u+\pi^2)}}=-\int_0^\infty {{\mbox{Li}_s\left(-{1 \over u}\right)} \over
{\ln^2 u+\pi^2}}du$$
$$=-\int_0^\infty {{\mbox{Li}_s(-v)dv} \over {v^2(\ln^2 v+\pi^2)}}.
\eqno(4.14)$$
(4.12) then immediately follows.  \qed

{\it Example}.  When $k=0$ in (4.11) and $s \to 1$,
$$\gamma_0=\gamma=\int_0^\infty {{\ln(1+v)} \over {v^2(\ln^2 v+\pi^2)}}dv, \eqno(4.15)$$
and this case is equivalent to (2.87) in \cite{coffey2009}.

Let $\Phi$ denote the Lerch zeta function, $\Phi(z,s,a)=\sum_{n=0}^\infty {z^n \over {(n+a)^s}}$, and analytically continued.  
This series holds for $s \in \mathbb{C}$ when $|z|<1$ and for Re $s>1$ when $|z|=1$.
The function $\Phi$ has an integral representation 
$$\Phi(z,s,a)={1 \over {\Gamma(s)}}\int_0^\infty {{t^{s-1}e^{-at}} \over {1-ze^{-t}}}dt=
{1 \over {\Gamma(s)}}\int_0^\infty {{t^{s-1}e^{-(a-1)t}} \over {e^t-z}}dt,$$
for Re $s>0$ when $|z|\leq 1$, $z \neq 1$, and for Re $s>1$ when $z=1$.
{\newline \bf Proposition 12}. 
$$\zeta(s,a)-{1 \over {(s-1)}}{1 \over a^{s-1}}=\int_0^\infty {{\Phi(-v,s,a)} \over {v(\ln^2 v
+\pi^2)}}dv.  \eqno(4.16)$$

{\it Proof} sketch.  Now
$$\int_0^1 u^{a-1}(-\ln u)^{s-1} (1-u)^{n-1}du={{\Gamma(s)} \over n}\sum_{k=1}^n {{(-1)^{k-1}k}
\over {(k+a-1)^s}} {n \choose k}.  \eqno(4.17)$$
Then (3.5) is used, following steps similar to the proof of Proposition 11, so that
$$\zeta(s,a)-{1 \over {(s-1)}}{1 \over a^{s-1}}=\sum_{k=0}^\infty \sum_{n=k}^\infty p_{n+2} {n \choose k} {{(-1)^k} \over {(k+a)^s}}.  \eqno(4.18)$$
Performing the sum on $k$ and changing variable in the integral gives (4.16).  \qed

\noindent
{\bf Corollary 7}.
$$\ln a-\psi(a)=\ln a+\gamma_0(a)=\int_0^\infty {{\Phi(-v,1,a)} \over {v(\ln^2 v+\pi^2)}}dv.  \eqno(4.19)$$

\medskip
\centerline{\bf Appendix}
\medskip

Here we show that for $n \in {\mathbb N}^+$
$${{(a)_j} \over {(a+n)_j}}={{(a)_n} \over {(n-1)!}}\sum_{k=0}^{n-1} {{n-1} \choose k}
{{(-1)^k} \over {(a+k)}}{{(a+k)_j} \over {(a+k+1)_j}}.  \eqno(A.1)$$

{\it Proof}.  We have the known decomposition
$${1 \over {x(x+1)\cdots(x+N)}}={1 \over {N!}}\sum_{k=0}^N {N \choose k}{{(-1)^k} \over
{(x+k)}}.  \eqno(A.2)$$
Then
$${{(a)_j} \over {(a+n)_j}}={{a(a+1)(a+2)\cdots (a+n-1)} \over {(a+j)(a+j+1)\cdots (a+j+n-1)}}$$
$$={{(a)_n} \over {(n-1)!}}\sum_{k=0}^{n-1} {{n-1} \choose k} {{(-1)^k} \over {(a+j+k)}}.
\eqno(A.3)$$
By noting that
$${{(a+k)_j} \over {(a+k+1)_j}}= {{(a+k)} \over {(a+k+j)}}, \eqno(A.4)$$
the result follows.  \qed

\pagebreak

\end{document}